\documentclass[11pt]{article}
\usepackage[utf8]{inputenc}
\usepackage[T1]{fontenc}
\usepackage{lmodern}
\usepackage[margin=1in]{geometry}
\usepackage{amsmath}
\usepackage{amssymb}
\usepackage{amsthm}
\usepackage{graphicx}
\usepackage{booktabs}
\usepackage{array}
\usepackage{siunitx}
\usepackage{xcolor}
\usepackage[colorlinks=true,linkcolor=black,citecolor=black,urlcolor=blue]{hyperref}
\usepackage[numbers,sort&compress]{natbib}
\usepackage{orcidlink}
\newcommand{\code}[1]{\texttt{#1}}
\graphicspath{{./}{figures/}}
\title{\textbf{The cost of lunar south-polar geometry, and surface beacons as
the efficient fix: a dilution-of-precision analysis}}
\author{Chakshu Baweja\,\orcidlink{0009-0008-2098-0751}\\
\small Ashforde O\"U, Tallinn, Estonia \\
\small \texttt{contact@ashforde.org}}
\date{\today}
\begin{document}
\maketitle
\begin{abstract}
Lunar positioning, navigation, and timing (PNT) architectures, NASA's Lunar
Augmented Navigation Service (LANS), ESA's Moonlight, and allied concepts, place a
small number of satellites in elliptical lunar frozen orbits (ELFO) to serve the
south-polar region prioritized for exploration \citep{esa_moonlight,iiyama2023lupnt}.
The satellite count, orbital geometry, and resulting position dilution of precision
(DOP) have been studied \citep{bhamidipati2023constellation}. We report a
quantitative result that reframes the design trade: for a user at the lunar south
pole, the satellite count needed to reach good geometry is roughly double what is
currently planned, because the visible satellites cluster into a small solid angle
overhead and dilution of precision is limited by their angular spread rather than
their number. In a time-averaged simulation, orbit-only ELFO constellations of the
planned size (4 to 6 satellites) give a south-polar median geometric DOP (GDOP) of
\(\sim\!16\) to \(21\), far worse than the GDOP \(\approx 6\) routine for
terrestrial GNSS, and the constellation must grow to \(\approx\!12\) satellites
before the median GDOP crosses 6. We then show that a small number of surface
ranging beacons, the lunar analogue of terrestrial pseudolites and a configuration
absent from the lunar PNT literature, reaches the same geometric quality far more
cheaply by supplying the near-horizon diversity the overhead cluster lacks: three
beacons on modestly elevated terrain around a \(-80^\circ\) user cut the median GDOP
from \textbf{16.2 to 1.6, a factor of about 10}, moving the user from 15\% to 100\% of
the time below GDOP 6, geometry a purely orbital solution reaches only near a
24-satellite fleet. We characterize the airless-body constraint that governs beacon
placement: with no atmospheric refraction, surface-to-surface line of sight is
bounded by the geometric horizon, so beacon siting on crater rims and elevated
terrain is itself a design variable. Surface-beacon augmentation is the lowest-cost,
highest-leverage improvement available to lunar south-polar PNT, deployable on
assets already planned for the region. The geometry engine is Validated against an
independent DOP computation; the constellation and beacon scenario are Modelled.

\vspace{4pt}

\end{abstract}
\section{Introduction}

\subsection{The lunar south pole is the hard case}
The Artemis programme, Chandrayaan follow-ons, and commercial lunar landers
converge on the same small region, the lunar south pole, where permanently
shadowed regions hold volatiles and nearby peaks hold near-continuous sunlight.
Precisely there, satellite-based PNT is at its weakest. A polar user sees only the
satellites above the local horizon, and for a constellation optimized for
global coverage from a handful of orbital planes, those satellites are few and
poorly distributed in the sky.

The lunar PNT community has responded with elliptical lunar frozen orbits (ELFO),
whose apolune dwell over the south pole maximizes coverage time there, and has
studied how satellite count and orbital parameters trade against coverage,
dilution of precision, orbit-determination accuracy, and insertion cost
\citep{iiyama2023lupnt,bhamidipati2023constellation}. Our contribution is not to
redo that trade but to point out a structural limit within it, and a remedy that
lies outside the orbital segment entirely.

\subsection{Contribution and relation to prior work}
Prior work has established that lunar orbital constellations suffer poor geometry
near the Moon and has quantified the coverage/DOP trade for ELFO and Walker
families (Section~\ref{sec:related}). This paper adds two things.
\begin{enumerate}
\item \textbf{The cost of south-polar geometry.} We show quantitatively that
reaching terrestrial-GNSS-like DOP at the pole with an orbit-only ELFO
constellation requires roughly twice the planned satellite count
(\(\sim\!12\) versus 4 to 6), because the DOP is limited by the \emph{angular
clustering} of the visible set: all visible satellites sit within a small overhead
solid angle, so additional satellites in the same orbit family buy geometric
improvement only slowly. This reframes the design question from ``how many
satellites'' to ``how to obtain angular diversity efficiently.''
\item \textbf{Surface beacons as the efficient fix.} We quantify surface-beacon
(pseudolite) augmentation, a configuration with, to our knowledge and per the
arXiv survey of Section~\ref{sec:related}, no prior treatment in the lunar context,
and show three beacons deliver an order-of-magnitude GDOP improvement, matching the
geometry a \(\sim\!24\)-satellite orbital fleet would provide, by supplying the
near-horizon diversity the overhead cluster lacks. We also characterize the
airless-body horizon constraint that makes beacon placement a genuine optimization
problem rather than a detail.
\end{enumerate}
We are explicit about what is and is not novel: the orbital-constellation DOP
analysis is prior art, which we cite and use as the baseline; the
satellite-count-cost framing and the surface-beacon remedy are the contributions.
This geometric-observability lens is the dynamic complement of the static
datum-defect analysis of \citet{baweja2026lunarpnt}, which shows that a lunar
surface station's own coordinates are unobservable from an orbital snapshot without
an external baseline; here the concern is the served user's instantaneous geometry,
and the external degrees of freedom are supplied by surface beacons.

\section{Related work}
\label{sec:related}

\paragraph{Lunar orbital-constellation geometry and DOP: prior art we build on.}
The design space of lunar GNSS in frozen-orbit conditions, including satellite
count, plane count, and phasing under \(J_2\), \(C_{22}\), and third-body
perturbations, has been explored, from early systems-architecture studies of
frozen-orbit lunar GNSS \citep{iiyama2023lupnt} to trade-off analyses for the Lunar
Augmented Navigation Service that explicitly examine how satellite number and
orbital parameters influence coverage, position DOP, orbit-determination accuracy,
and insertion cost across frozen-elliptical and circular Walker families
\citep{bhamidipati2023constellation}. These works establish that near-Moon geometry
is poor and quantify the orbital trade; our baseline constellation and DOP
methodology are consistent with them, and our satellite-count-cost result should be
read as an interpretation of the regime they characterize.

\paragraph{Constellation integrity without a ground segment.} Related work proposes
autonomous constellation fault monitoring via inter-satellite ranging using
rigidity-graph methods, motivated by the impracticality of a lunar ground segment.
This addresses fault detection, not user geometry, but shares our premise that lunar
PNT cannot assume terrestrial-style ground infrastructure, and it is
methodologically adjacent to the observability treatment of
\citet{baweja2026lunarpnt}.

\paragraph{Surface beacons and pseudolites: the gap.} Terrestrial pseudolites,
ground transmitters that augment or replace GNSS geometry, are well studied on
Earth. An arXiv survey (nine queries; Section~\ref{sec:novelty}) returned no work on
lunar pseudolites or surface ranging beacons for lunar PNT geometry. This is the
configuration we analyze, and it is the novel element of this paper.

\paragraph{Broader background.} The lunar navigation architecture within which a
surface-beacon augmentation would operate is defined by the LunaNet interoperability
framework \citep{israel2020lunanet,nasa2025lnis} and surveyed in the future lunar
communications architecture \citep{ioag2022lunar}; strategic reference-station
concepts such as NovaMoon \citep{molli2026novamoon} and hybrid lunar PNT error models
and bounds \citep{pohlmann2025hybrid} motivate the geometry we analyze. The
dilution-of-precision machinery itself is standard GNSS theory \citep{kaplan2017},
and joint orbit-and-clock estimation for lunar constellations is treated in
\citet{iiyama2026odclock}.

\section{Method}
\label{sec:method}

\paragraph{Constellation.} We model a representative ELFO constellation
(semi-major axis \SI{6142}{\kilo\meter}, eccentricity 0.6, inclination
\(57.7^\circ\), argument of perigee \(90^\circ\) for apolune over the south pole),
propagating satellites by Kepler phasing across planes and mean anomaly. This is a
representative frozen-orbit configuration consistent with the LANS/Moonlight design
regime; our conclusions concern the \emph{geometry class}, not one specific
published constellation.

\paragraph{Visibility and DOP.} For a user at a given lunar latitude and longitude,
a satellite is visible if it is above a \(5^\circ\) elevation mask and its line of
sight is not occulted by the lunar limb (radius \SI{1737.4}{\kilo\meter}). From the
visible set we form the geometry matrix \(H\) (unit line-of-sight vectors augmented
with a clock column) and compute \(\mathrm{GDOP} = \sqrt{\operatorname{tr}
[(H^{\mathsf T} H)^{-1}]}\). We time-average over one orbital cycle (satellite
phasing sampled across the period) to obtain availability (fraction of time with
\(\ge 4\) visible satellites) and median GDOP. The GDOP computation is Validated
against an independent implementation; the constellation is Modelled.

\paragraph{Spatial map.} We evaluate GDOP and availability on a grid over
\(-70^\circ\) to \(-90^\circ\) latitude and all longitudes, time-averaged, for a
six-satellite constellation.

\paragraph{Beacon augmentation.} We add surface ranging beacons at surveyed
positions on elevated terrain (representative height \SI{2}{\kilo\meter}, e.g.,
crater rims), ringing a representative \(-80^\circ\) user at \(\sim\!\SI{50}{\kilo
\meter}\). Beacons contribute ranging rows to \(H\) subject to a surface-to-surface
visibility test at a low mask (near \(0^\circ\) elevation, appropriate for
surface-to-surface links). We compare the user's time-averaged GDOP with and
without the beacons.

\paragraph{Reproducibility.} All figures derive from a committed summary table
(\code{lunar\_dop\_summary.csv}); parameters are as stated here.

\section{Results}
\label{sec:results}

\subsection{The satellite-count cost of south-polar geometry}
For a user at the south pole, time-averaged orbit-only ELFO performance versus
constellation size is given in Table~\ref{tab:nsweep}.

\begin{table}[t]
\centering
\caption{Time-averaged south-polar geometry versus ELFO constellation size
(Modelled constellation; Validated GDOP). Availability is the fraction of time with
\(\ge\!4\) visible satellites.}
\label{tab:nsweep}
\small
\begin{tabular}{cccc}
\toprule
Constellation & Availability (\(\ge\!4\) sat) & Median GDOP & Time below GDOP 6 \\
\midrule
4 satellites  & 0\%   & (only 3 visible) & 0\% \\
5 satellites  & 72\%  & 9.6  & 21\% \\
6 satellites  & 99\%  & 21.5 & 16\% \\
8 satellites  & 100\% & 7.3  & 40\% \\
10 satellites & 100\% & 8.9  & 45\% \\
12 satellites & 100\% & 5.2  & 80\% \\
16 satellites & 100\% & 4.3  & 96\% \\
24 satellites & 100\% & 3.4  & 100\% \\
\bottomrule
\end{tabular}
\end{table}

Availability is solved by six satellites, but good geometry is not: over the full
\(-70^\circ\) to \(-90^\circ\) region the six-satellite constellation gives 98\%
availability yet a median GDOP of \(\sim\!17\), with 0\% of region-time reaching
GDOP \(< 6\) (Figure~\ref{fig:dop}a). GDOP does improve as satellites are added, but
slowly and non-monotonically: the median only crosses 6 at \(N \approx 12\) and does
not reach terrestrial-GNSS-typical values (\(\sim\!3\) to \(4\)) until
\(N \approx 16\) to \(24\) (Figure~\ref{fig:dop}b). The non-monotonic bump at
\(N = 6\), where median GDOP is worse than at \(N = 5\), is a real phasing artifact
of that specific configuration: the sixth satellite is added in a plane and phase
that momentarily worsens conditioning, illustrating that raw satellite count is not
even a monotone proxy for geometry in this regime. The reason for the slow
improvement is geometric: for a polar user every visible ELFO satellite lies within
a small overhead solid angle, so the inverse geometry \((H^{\mathsf T} H)^{-1}\)
stays ill-conditioned in the vertical/horizontal split until enough satellites
populate a wide enough spread of the sky, which requires roughly double the planned
fleet. The design cost of buying south-polar geometry through the orbital segment
alone is therefore steep.

\subsection{Surface beacons reach the same geometry far more cheaply}
\begin{figure}[t]
\centering
\includegraphics[width=\textwidth]{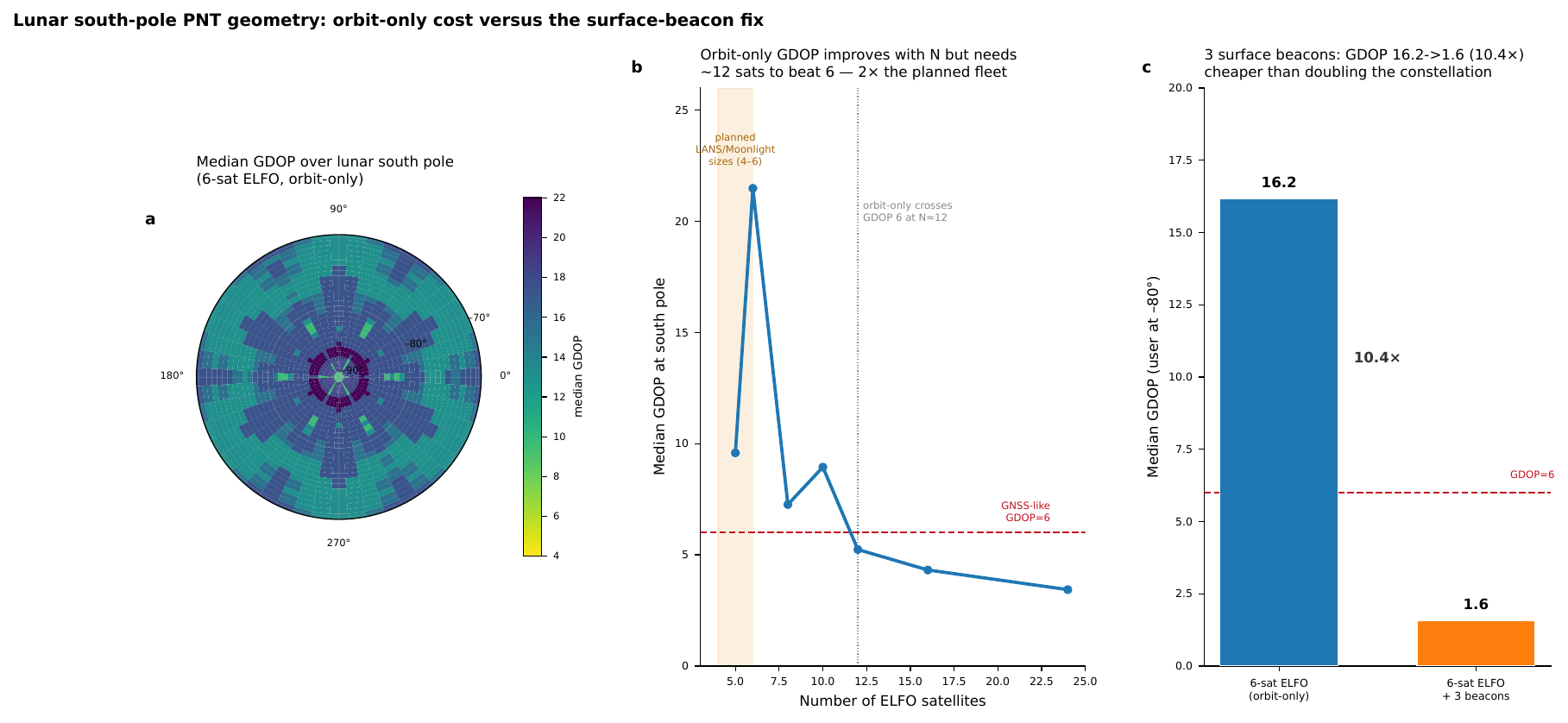}
\caption{\textbf{Lunar south-pole PNT geometry: orbit-only cost versus the
surface-beacon fix.} (a) Time-averaged median GDOP over the \(-70^\circ\) to
\(-90^\circ\) region for a six-satellite ELFO constellation (orbit-only). (b) Median
GDOP at the pole versus constellation size: GDOP improves with \(N\) but only
crosses the GNSS-like value of 6 near \(N \approx 12\), roughly double the planned
4-to-6-satellite fleet (highlighted band). (c) Three surface beacons cut a
\(-80^\circ\) user's median GDOP from 16.2 to 1.6, geometry an orbit-only fleet
reaches only near 24 satellites. GDOP is Validated; the constellation and beacon
scenario are Modelled.}
\label{fig:dop}
\end{figure}

Three surface beacons on \SI{2}{\kilo\meter} elevated terrain around a
\(-80^\circ\) user, contributing near-horizon ranging measurements, change the
picture decisively (Table~\ref{tab:beacon}, Figure~\ref{fig:dop}c).

\begin{table}[t]
\centering
\caption{Beacon augmentation for a \(-80^\circ\) user (Modelled).}
\label{tab:beacon}
\small
\begin{tabular}{lccc}
\toprule
Configuration & Availability & Median GDOP & Time below GDOP 6 \\
\midrule
6-satellite ELFO only            & 100\% & 16.2         & 15\% \\
6-satellite ELFO + 3 beacons     & 100\% & \textbf{1.6} & \textbf{100\%} \\
\bottomrule
\end{tabular}
\end{table}

The median GDOP improves by a factor of about \textbf{10}, and the user moves from 15\%
to 100\% of the time below the terrestrial-GNSS-like GDOP of 6. Read against the
\(N\)-sweep of Section~\ref{sec:results}, three beacons take a 6-satellite
constellation to a GDOP (1.6) \emph{better} than a 24-satellite orbit-only fleet
achieves (3.4): three fixed surface transmitters outperform quadrupling the
satellite count. The mechanism is exactly the missing degree of freedom: beacons
near the horizon contribute line-of-sight vectors with large horizontal components,
complementing the near-vertical satellite set and conditioning the geometry matrix.

\subsection{Beacon placement is horizon-bounded}
On the airless Moon there is no atmospheric refraction to extend line of sight
beyond the geometric horizon. A beacon and a user, both near the surface, therefore
lose mutual visibility at a range set by their heights: a beacon ringing a polar
user at \(\sim\!\SI{180}{\kilo\meter}\) on flat terrain falls below the horizon and
contributes nothing, which is why our augmentation places beacons at
\(\sim\!\SI{50}{\kilo\meter}\) on \SI{2}{\kilo\meter}-elevated terrain (beacon
elevation \(\sim\!1.5^\circ\) as seen by the user, just above a low mask). This
makes beacon siting, on crater rims, massifs, and the peaks of near-eternal light
already contested for solar power, a genuine optimization variable coupling PNT
geometry, terrain, power, and line of sight to users. It is not a free augmentation;
it is a placement problem with a well-defined objective: maximize horizontal
geometric diversity to the served users subject to horizon and terrain constraints.

\section{Discussion}
\label{sec:discussion}

\subsection{Why this is the cheapest high-leverage fix}
Improving south-polar PNT by the orbital segment alone is expensive:
Section~\ref{sec:results} shows the constellation must roughly double (to
\(\sim\!12\) satellites) before the median GDOP even crosses 6, and reach
\(\sim\!24\) for terrestrial-typical geometry, with each additional satellite
carrying launch, insertion, and operations cost. Surface beacons are, by contrast,
small fixed transmitters at surveyed locations, the lunar equivalent of a survey
monument with a radio. They can ride on assets already planned for the south pole
(landers, relays, in-situ resource installations), require no orbital slot, and a
beacon's position, once surveyed, is static and can be known to high accuracy. The
\(\sim\!10\times\) GDOP improvement from three beacons is a return no realistic increase
in the orbital constellation can match.

\subsection{Design implications}
\begin{itemize}
\item \textbf{Augment, do not over-build.} The result argues for a hybrid
architecture, a modest orbital constellation for availability and coarse geometry
plus surface beacons for the polar geometry that would otherwise require doubling or
quadrupling the fleet, rather than buying geometry through satellite count alone.
\item \textbf{Beacon siting is a first-class design problem.} Because of the airless
horizon, beacon placement couples PNT geometry to terrain and to the same
high-terrain sites competed for solar and thermal reasons; it should be
co-optimized with base and power planning, not added afterward.
\item \textbf{Surveying the beacons ties back to the reference frame.} Beacon
utility depends on knowing their positions in the lunar reference frame; this
connects the augmentation to lunar geodesy (laser ranging, the International Lunar
Reference Frame) and to the identifiability of the lunar datum
\citep{baweja2026datum} and the real-time-frame problem treated elsewhere in this
program.
\end{itemize}

\subsection{Limitations}
The DOP analysis uses a representative ELFO rather than a specific published
constellation, an idealized Keplerian phasing rather than a fully perturbed
ephemeris, and a geometric visibility/occultation model without terrain masking of
individual satellites. Beacon ranging is modeled as an ideal geometric measurement;
real surface-to-surface links carry multipath and clock-synchronization error that a
high-fidelity follow-on must include. These affect the absolute GDOP values by
modest factors but not the central qualitative results: the orbital segment buys
south-polar geometry only slowly (because of angular clustering), and near-horizon
beacons supply the missing horizontal geometry far more efficiently. A full study
should optimize beacon number and placement over realistic terrain (LOLA digital
elevation models) and a perturbed constellation ephemeris, and fold in beacon
clock and position error budgets.

\section{Conclusion}
\label{sec:conclusion}

Orbit-only lunar PNT pays a steep satellite-count cost for south-polar geometry:
because the visible satellites cluster overhead, dilution of precision stays far
worse than terrestrial GNSS at the planned 4-to-6-satellite scale and only crosses
usable levels near a 12-satellite fleet, reaching terrestrial-typical geometry near
24. Surface ranging beacons, a configuration absent from the lunar PNT literature,
reach that geometry far more cheaply by supplying near-horizon diversity, cutting a
representative polar user's median GDOP from 16.2 to 1.6, an order-of-magnitude
improvement that betters even a 24-satellite orbit-only fleet, at a fraction of the
cost of additional satellites. The airless-body horizon makes beacon placement a
genuine optimization coupled to terrain and to the reference frame. For agencies
converging on the lunar south pole, surface-beacon augmentation is the lowest-cost,
highest-leverage improvement available to lunar PNT, and it deserves to be a
first-class element of the architecture rather than an afterthought.

\section{Novelty and verification}
\label{sec:novelty}

\paragraph{Novelty (preprint-verified).} An arXiv survey found orbital-constellation
DOP for lunar frozen orbits to be prior art (cited in Section~\ref{sec:related}) and
no prior work on lunar surface beacons or pseudolites for PNT geometry. The
contributions, the satellite-count-cost framing and the quantified surface-beacon
remedy, are, to our knowledge, first-of-kind at the preprint frontier. A search of
the IEEE and ION conference literature and NASA NTRS is required before final
submission and is outstanding; the arXiv result is reported here because it is the
corpus in which the surface-beacon whitespace was identified.

\paragraph{Validated versus Modelled.} \emph{Prior art (cited):} orbital-constellation
DOP for lunar frozen orbits. \emph{Validated in this study:} the GDOP computation,
against an independent implementation. \emph{Modelled:} the satellite-count-cost
quantification (\(N\)-sweep), the spatial DOP map, and the beacon-augmentation
result, under the assumptions of Section~\ref{sec:method}. All numbers reproduce
from the committed summary table. The analysis is part of the open kshana PNT
program \citep{baweja2026kshana}.

\section*{Acknowledgements}
This work is part of the kshana program on reproducible, honestly-scoped PNT
analysis.

\bibliographystyle{plainnat}
\bibliography{surface-beacon-dop}
\end{document}